\documentclass[aip,apm,reprint,amsmath,amssymb]{revtex4-1}

\usepackage{graphicx}
\usepackage{dcolumn}
\usepackage{bm}
\begin{document}
\title{Excited exciton and biexciton localised states in a single quantum ring}
\author{H. D. Kim}
\affiliation{Clarendon Laboratory, University of Oxford, Oxford,
OX1 3PU, U.K.}
\author{K. Kyhm}\email{kskyhm@pusan.ac.kr}
\affiliation{Graduate School of Cogno-Mechatronics, Department of Physics
Education, Pusan Nat'l University, Busan, 609-735, South Korea.}
\affiliation{LNCMI and Institut N\'{e}el, CNRS, rue des Martyrs
38054, Grenoble, France.}
\author{R. A. Taylor}\email{r.taylor1@physics.ox.ac.uk}
\affiliation{Clarendon Laboratory, University of Oxford, Oxford,
OX1 3PU, U.K.}
\author{A. A. L.  Nicolet}
\author{M. Potemski}
\author{G. Nogues}
\affiliation{LNCMI and Institut N\'{e}el, CNRS, rue des Martyrs
38054, Grenoble, France.}
\author{K. C. Je}
\affiliation{Department of Physics, Anyang University, Anyang 430-714, South Korea.}
\author{E. H. Lee}
\author{J. D. Song}
\affiliation{Nano-Photonics Research Center, KIST, Seoul 136-791, South Korea.}
\date{\today}
\begin{abstract}
We observe excited exciton and biexciton states of localised
excitons in an anisotropic quantum ring, where large polarisation
asymmetry supports the presence of a crescent-like localised structure.
We also find that saturation of the localised ground state exciton with
increasing excitation can be attributed to relatively fast
dissociation of biexcitons ($\sim430\,$ps) compared to slow
relaxation from the excited state to the ground state
($\sim1000\,$ps). As no significant excitonic Aharonov-Bohm
oscillations occur up to 14 T, we conclude that phase coherence around the
rim is inhibited as a consequence of height anisotropy in the quantum ring.
\end{abstract}
\maketitle \indent Thanks to droplet epitaxy
\cite{Mano,Warburton,Spain}, quantum rings (QRs) have become an
alternative means of studying the Aharonov-Bohm (AB) effect, which manifests itself by
a persistent current\cite{Holland,pcurrent} or an
oscillation of the conductance\cite{ABnature} and emission
energy\cite{Teodoro,Govorov1,Govorov2,Ding,ZnTetypeII,Ulloa} for
an external magnetic field. A delocalised wavefunction around the
rim is a prerequisite for the AB effect, but the presence of
localised states has been reported in a volcano-like QR arising from
height anisotropy\cite{Holland,Taiwan,Kim}, whereby the phase
coherence around the rim can be inhibited. The degree of localisation
seems to depend upon the anisotropy; when this is
negligible\cite{Ding}, excitonic AB oscillations begin at low
magnetic fields. In the intermediate range of localisation,
excitons are localised separately near the two highest rims in a
QR, but the two excitons can be paired in a biexciton with
sub-meV binding energy\cite{Taiwan}. In the case of strong
localisation in a QR\cite{Holland}, no persistent current emerges
unless a large magnetic field is applied. Optical AB
oscillations in a QR have not been considered in the context of
the localisation induced by rim height-anisotropy. This may
explain why some experiments failed to observe excitonic AB
oscillations in a QR\cite{Taiwan, noAB}. In this work, we confirm
the presence of a strongly localised state in a volcano-like QR in
terms of an absence of excitonic AB oscillations up to 14$\,$T, with a small
diamagnetic coefficient the polar nature and large polarisation asymmetry of biexcitons and excited excitons. 
These arise as a consequence of the
small lateral extension and asymmetry of the strongly localised
crescent-like structure in the QR.\\
\indent GaAs rings\cite{Mano} were grown on an n-doped GaAs
(001) substrate using a molecular beam epitaxy system with an ion
getter pump, and capped with 60 nm-thick Al$_{0.3}$Ga$_{0.7}$As
and 3 nm-thick GaAs for optical measurement. The sample was
excited by frequency-doubled ($400\,$nm) Ti:sapphire laser pulses
($120\,$fs duration at $80\,$-MHz repetition rate). The
photoluminescence (PL) of a single QR was collected at $4.2\,$K
using a confocal arrangement, and a time-correlated single photon
counting system was used to obtain the time-resolved PL (TRPL).
Magneto-PL from a single QR was also performed in a resistive DC magnet (52 mm-bore diameter) under continuous-wave Ar$^+$-ion laser excitation (488 nm).\\
\begin{figure}
\includegraphics[width=7 cm]{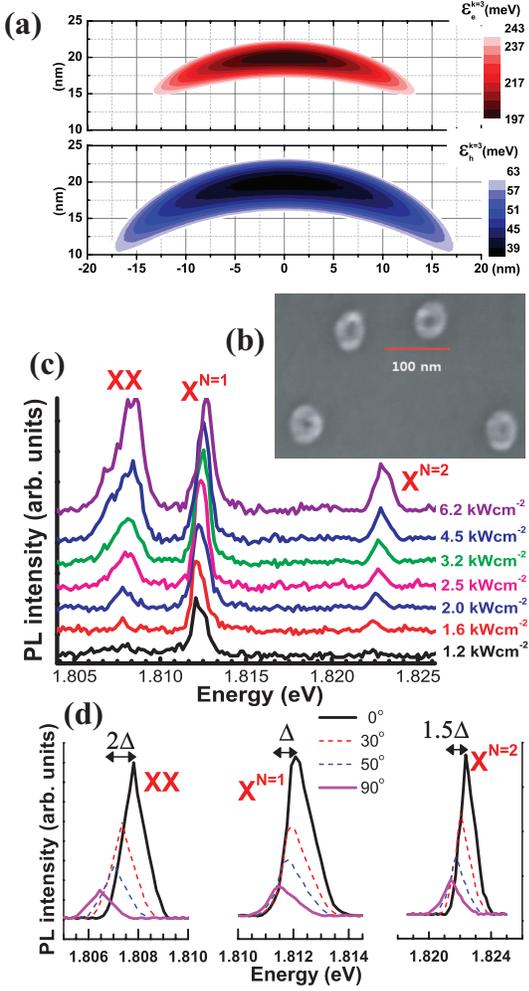}
\caption{(a) Electrons and holes are localised in crescent-like
adiabatic potentials for a vertical quantum number $k=3$
($\varepsilon^{k=3}_{\rm{e,h}}$). (b) A single ring structure
observed by FESEM. Biexciton (XX) and excited state exciton
(X$^{N=2}$) emission appear near ground exciton (X$^{N=1}$) for
increasing excitation power (c), among which analyzer angle dependence
of the PL spectrum are compared under an excitation power of
$2\,$kWcm$^{-2}$ (d).}
\end{figure}
\indent Although ring structures were observed in a field emission
scanning electron microscope (FESEM) image of the uncapped sample
(Fig.1(b)), it is known that a volcano-like anisotropic morphology
is present in the QR\cite{Holland}. The height anisotropy gives rise to a
crescent-shaped adiabatic potential and localisation
of the potential becomes significant as the vertical quantum
number is increased\cite{Kim}. For a QR with $\sim20\,$nm-radius and
$\sim10\,$nm-height, localised states for a vertical quantum number of $k=3$
can be seen at an energy close to the barrier
(Al$_{0.3}$Ga$_{0.7}$As) bandgap. As shown in Fig.1(a), localised
adiabatic potentials ($\varepsilon^{k=3}_{\rm{e,h}}$) of the
electron and the hole for a vertical quantum number of $k=3$ have
a crescent-like shape. An estimated ground state energy for the electron ($241\,$meV) and the hole ($60\,$meV) predicts a PL energy for the ground state exciton
X$^{N=1}$ of $\sim1.812\,$eV in excellent agreement with
the PL spectrum in Fig.1(c). It should be noted that
the wavefunction contour areas should be larger than the localised
potentials due to tunnelling effects.\\
\indent The blueshift in the PL from X$^{N=1}$ with increasing
excitation power (Fig.1(c)) suggests the presence of fine structure states,
where a sequential state-filling gives rise to the observed
blueshift. As a consequence of asymmetry of the crescent-like
adiabatic potential in Fig.1(a), the fine structure states of X$^{N=1}$
within the $\sim1.5\,$meV PL linewidth can be resolved by
measuring the PL at different linear polariser angles (Fig.1(d)).
As the excitation power ($I_{\rm{ex}}$) is increased, two
additional PL peaks emerge at low ($1.808\,$eV) and high energy
($1.822\,$eV) with respect to X$^{N=1}$ ($1.812\,$eV). The
superlinear increase of the PL intensity ($\sim
I_{\rm{ex}}^{\alpha}$) was characterized in terms of the power
factor ($\alpha$) by integrating the PL spectrum.
$\alpha\sim2.3\pm0.1$ and $\sim1.5\pm0.1$ were obtained for the
low and the high energy peaks compared with $\alpha\sim0.9\pm0.1$
measured for X$^{N=1}$ before saturation of the PL intensity.
Therefore, the two additional peaks can be attributed to biexciton
states (XX) and excited exciton states (X$^{N=2}$) of the
localised X$^{N=1}$, respectively, where the radial quantum number
$N$ denotes the states defined in the adiabatic potential
($\varepsilon^{k=3}_{\rm{e,h}}$). Both wavefunctions for XX and
X$^{N=2}$ are possibly more extended and asymmetric than that of
X$^{N=1}$. However, as X$^{N=2}$ is located $\sim10\,$meV above X$^{N=1}$,
the range of X$^{N=2}$ is not extended significantly ($\sim$few nm) in the contour areas of $\varepsilon^{k=3}_{\rm{e}}$ and $\varepsilon^{k=3}_{\rm{h}}$,
but a node of the wavefunction must exist in the middle of the crescent structure similar to $p$-orbitals.
This possibly results in a large polarisation asymmetry and and points to the polar nature of the wavefunction. Since XX consists of two X$^{N=1}$s,
the area could be nearly doubled, but shrinks due to bonding. As the observed binding energy of the XX is large ($\sim4\,$meV),
this is the case of a strongly localised XX in a crescent-like structure rather than a pair of two different excitons,
which are located separately at different rims\cite{Taiwan}. Again, the asymmetry of the local structure gives rise to a strong polarisation dependence,
where the emission energy difference of XX for perpendicular polarisations is twice ($2\Delta$) that of X$^{N=1}$ ($\Delta\sim0.8\,$mev) due to selection rule breaking\cite{Kim}. On the other hand, the emission energy difference of X$^{N=2}$ for perpendicular polarisations is $1.5\Delta$. Although X$^{N=2}$ possibly has a relatively smaller wavefuction area than XX, the wavefunction shape should be very asymmetric, i.e., a node is present as in $p$-orbitals and is confined in the crescent-like structure. Therefore, the emission energy difference in the level spacing between the fine structure states of X$^{N=2}$ is relatively large compared to that of X$^{N=1}$.\\
\begin{figure}[t]
\includegraphics[width=6.5 cm]{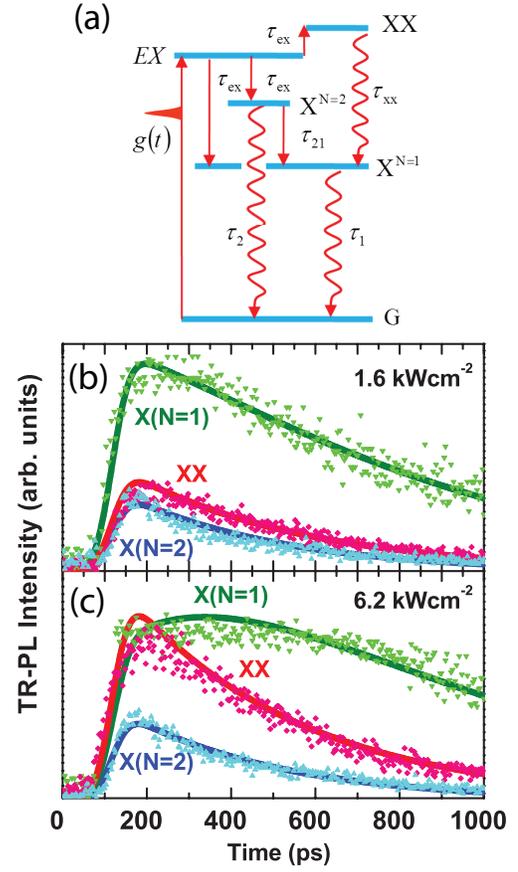}
\caption{(a) Schematic diagram of the transitions between the
energy levels of the excited (EX) state, X$^{N=1}$, X$^{N=2}$, and
XX, where each TRPL spectrum was measured for an excitation power
of $1.6\,$kWcm$^{-2}$ (b) and $6.2\,$kWcm$^{-2}$ (c). The
calculations (solid lines) are in agreement with the experimental
results (dots).}
\end{figure}
\indent As shown schematically in Fig.2(a), the underlying
dynamics of TRPL can be modelled by using coupled rate equations
for X$^{N=1}$, X $^{N=2}$, and XX:
\begin{eqnarray}
\frac{{\rm d}N_{\rm ex}}{{\rm d}t}&=&g(t)-\frac{N_{\rm ex}}{\tau_{\rm ex}},\\
\frac{{\rm d}N_{\rm xx}}{{\rm d}t}&=&\gamma\frac{N_{\rm ex}}{\tau_{\rm ex}}-\frac{N_{\rm xx}}{\tau_{\rm xx}},\\
\frac{{\rm d}N_{\rm x_{2}}}{{\rm d}t}&=&\alpha\frac{N_{\rm ex}}{\tau_{\rm ex}}-\frac{N_{\rm x_{2}}}{\tau_2}-\frac{N_{\rm x_{2}}}{\tau_{21}},\\
\frac{{\rm d}N_{\rm x_{1}}}{{\rm d}t}&=&\beta\frac{N_{\rm
ex}}{\tau_{\rm ex}}+\frac{N_{\rm xx}}{\tau_{\rm xx}}+\frac{N_{\rm
x_{2}}}{\tau_{21}}-\frac{N_{\rm x_{1}}}{\tau_1}.
\end{eqnarray}
\indent Initially, carriers are generated in the excited state
(denoted by ex) from the ground state by injection of a laser
pulse ($g(t)$), which then relax quickly to X$^{N=1}$, X $^{N=2}$,
and XX in a relaxation time $\tau_{\rm ex}$. Because of the
limited time-resolution of our TCSPC system ($\sim50\,$ps), we
cannot determine $\tau_{\rm ex}$, therefore the initial population
of the X$^{N=1}$ ($N_{\rm x_{1}}$), X$^{N=2}$ ($N_{\rm x_{2}}$),
and XX ($N_{\rm xx}$) are given by multiplying the
intra-relaxation rate ($1/\tau_{\rm ex}$) by the weight factors
($\beta$, $\alpha$, and $\gamma$), respectively, i.e., the initial
carriers are distributed among X$^{N=1}$, X$^{N=2}$, and XX such
that $\alpha+\beta+\gamma=1$. While the XX  dissociates into
X$^{N=1}$ by radiative recombination during $\tau_{\rm xx}$,
X$^{N=2}$ can either relax into X$^{N=1}$ or decay radiatively to
the ground state. As a result, $N_{\rm x_{1}}$ can be enhanced by
the intra-relaxation of X$^{N=2}$ or the radiative dissociation of
XX. At an excitation power near the XX onset ($1.6\,$kWcm$^{-2}$)
(Fig.2(b)), the PL decay time of X$^{N=2}$ is comparable to that
of XX despite the presence of the two kinds of processes in the X
$^{N=2}$. Therefore, the intra-relaxation between X$^{N=2}$ and
X$^{N=1}$ is possibly slow, otherwise, $N_{\rm x_{2}}$ would show
a faster decay than that of XX. When the initial weight factors
($\alpha=0.2$ and $\gamma=0.26$) of X$^{N=2}$ and XX are used,
optimum time constants for the intra-relaxation
($\tau_{21}=1000\pm20\,$ps) and radiative recombination times of
X$^{N=2}$ ($\tau_{2}=600\pm20\,$ps), X$^{N=1}$
($\tau_{1}=500\pm15\,$ps), and XX ($\tau_{\rm xx}=450\pm10\,$ps)
are obtained. These results confirm that the $N_{\rm x_{1}}$
enhancement is dominated by $N_{\rm xx}$. A similar result has
been obtained at the higher excitation power of
$6.2\,$kWcm$^{-2}$, where the PL intensity of X$^{N=1}$ is
saturated. When compared with the data for $1.6\,$kWcm$^{-2}$, the
initial TRPL intensity of both X$^{N=2}$ and XX is increased with
respect to that of X$^{N=1}$, but no significant change in the
decay dynamics of X$^{N=1}$ and XX is seen (a slight change in the
error bar; $\tau_{\rm xx}=430\pm10\,$ps is observed). We found
again that the $N_{\rm x_{1}}$ enhancement is attributed to
$N_{\rm xx}$; the weight factor increase of XX ($\gamma=0.44)$ is
critical whilst other constants barely change within the fitting
error. Those results suggest that the relatively fast dissociation
of XX fills up the X$^{N=1}$ states, then the occupied states of
X$^{N=1}$ block the intra-relaxation between X$^{N=2}$ and
X$^{N=1}$. It is interesting that the ratio of PL decay rate
between XX and X$^{N=2}$
($\tau^{-1}_{\rm{xx}}/\tau^{-1}_{2}\sim1.33$) is in remarkable
agreement with the ratio of the energy difference for
perpendicular polarisations ($2\Delta/1.5\Delta\sim1.33$).\\
\begin{figure}[t]
\includegraphics[width=7 cm] {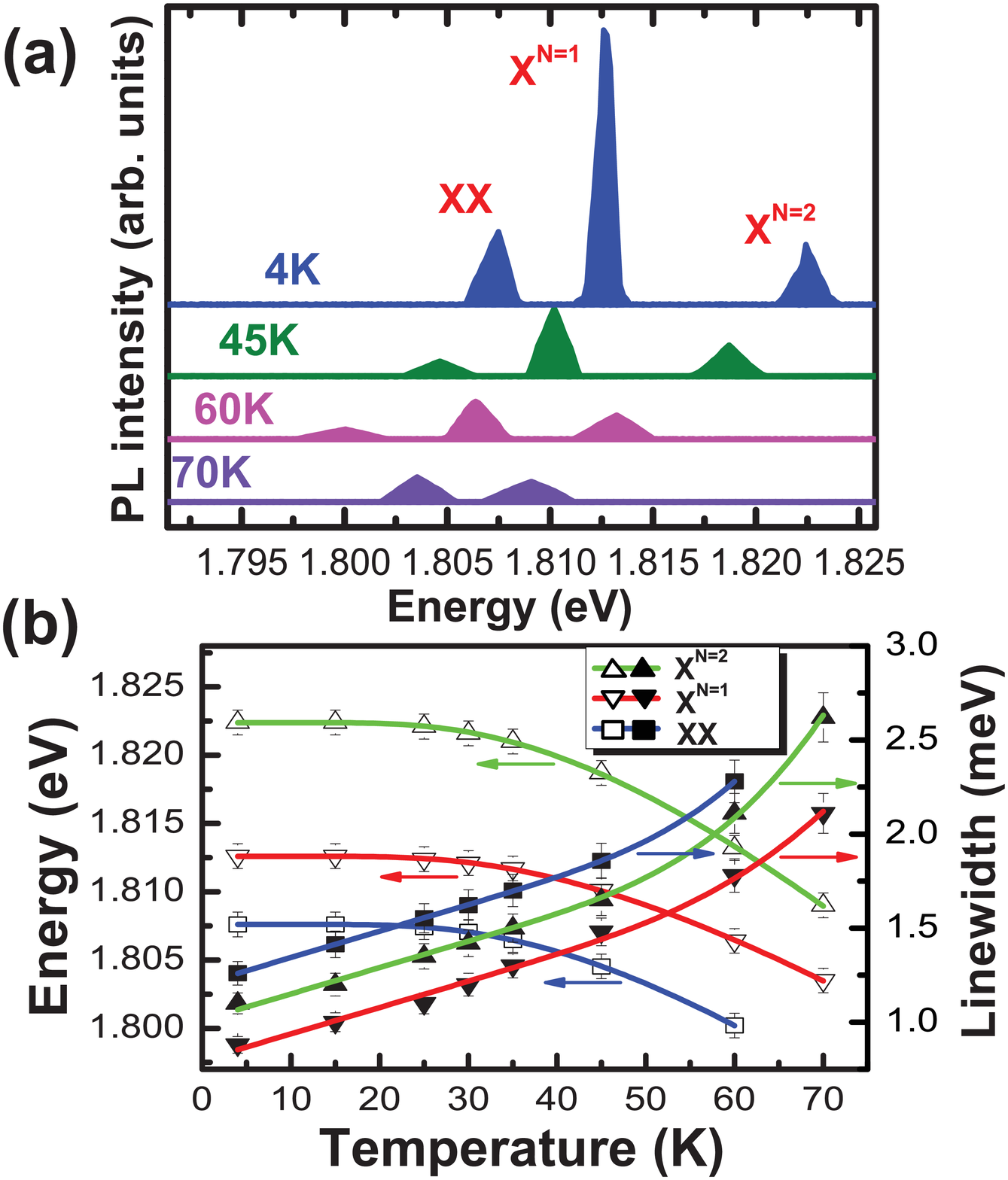}
\caption{ (a) Temperature dependence of the PL spectrum for X$^{N=1}$, X$^{N=2}$, and XX with at an excitation power of $2\,$kWcm$^{-2}$. (b) Temperature dependence of the PL energy (open) and the linewidth (filled) for X$^{N=1}$, X$^{N=2}$, and XX compared with theory (line).}
\end{figure}
\indent Additionally, asymmetry of the crescent-like local
structure possibly gives rise to a large polarisation, whereby the optical-phonon
interaction can be enhanced. This issue can be studied in terms of
the broadening of the PL linewidth as a function of temperature.
As shown in Fig.3(a), the PL intensities for all three states
decrease with increasing temperature, but the PL intensity of
X$^{N=2}$ becomes relatively more dominant due to thermal
excitation. Additionally, both X$^{N=2}$ and XX are also dominant
compared to X$^{N=1}$ in the spectral redshift and the linewidth
broadening as temperature is increased (Fig.3(b)). Given
longitudinal optical (LO) phonon energy in GaAs ($\hbar\omega_{\rm
LO}\sim36\,$meV), the temperature dependent linewidth
($\Gamma(T)$) (Fig.3(b)) is described by
$\Gamma(T)=\Gamma^{'}+c_{\rm{ac}}T+c_{\rm
opt}[\exp(\hbar\omega_{\rm{LO}}/k_{\rm{B}}T)-1]^{-1}$, where the
temperature-independent linewidth ($\Gamma^{'}=1.51\,\Gamma_{0}
(\rm XX), 1.27\,\Gamma_{0} (\rm {X}^{ N=2})$) and the coefficient
of optical phonon scattering ($c_{\rm opt}=1.81\alpha_{0}(\rm XX),
1.89\alpha_{0}$ (X$^{N=2}$)) for XX and X$^{N=2}$ were both found
to be large compared to those for X$^{N=1}$
($\Gamma_{0}\sim0.79\,$meV, $\alpha_{0}\sim132\,$meV), but no
significant difference was found in the coefficient of acoustic
phonon scattering ($c_{\rm{ac}}\sim14\,\mu$eV/K) amongst all three
lines X$^{N=1}$, X$^{N=2}$, and XX. The temperature dependence of
the energy gap shift ($\Delta E(T)$), which is dominated by phonon
interactions, is associated with the Bose-Einstein distribution as
$\Delta E(T)=-\gamma_{\rm{ph}}[\exp(T_{\rm{ph}}/T)-1]^{-1}$, where
the coupling energy to the phonon bath for XX and X$^{N=2}$
($\gamma_{\rm ph}=1.26\,\gamma_{0}$ (XX), $1.50\,\gamma_{0}$ (X$^{
N=2}$)) was found also to be large compared to that of X$^{N=1}$$
(\gamma_{0}\sim34\,$ meV) with an optimum effective phonon
temperature of $T_{\rm ph}\sim150\,$K. As expected, these results
confirm the stronger phonon interaction of XX and X$^{N=2}$
compared to that of X$^{N=1}$. It is also interesting that
X$^{N=2}$ has a slightly larger $c_{\rm opt}$ and $\gamma_{\rm
ph}$ than XX. This supports the conjecture of a large polarisation in the X$^{N=2}$ state due to the asymmetric $p$-orbital shape in the crescent-like local structure.\\
\begin{figure}[t]
\includegraphics[width=7 cm] {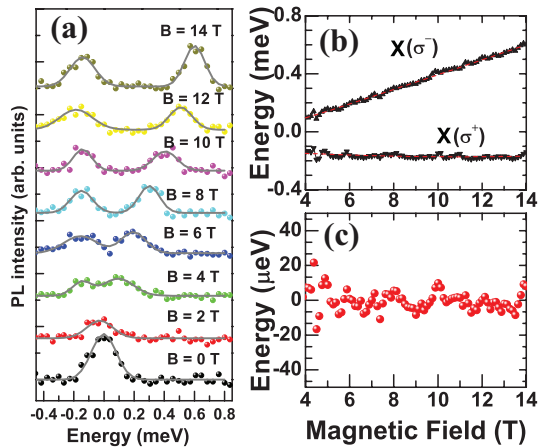}
\caption{The PL spectrum (a) and fitted peak energy (b) of
localised exciton Zeeman doublet with increasing an external
magnetic field, where the energy oscillation is not significant
with an error range of 20$\,\mu$eV after removing Zeeman splitting
and diamagnetic shift (c).}
\end{figure}
\indent As X$^{N=1}$, X$^{N=2}$, and XX are all localised,
delocalisation around the whole rim in a QR is unlikely. However,
provided that an external magnetic field ($B$) is strong enough to
overcome the energy barrier between the two separate localised
states in an anisotropic QR, a phase coherent delocalisation can be
induced; as persistent current\cite{Holland} was induced in an
anisotropic QR beyond $\sim14\,$T, it is challenging to observe an
induced excitonic AB oscillation in an anisotropic QR beyond a
threshold magnetic field. We measured magneto-PL of localised
excitons (Fig.4(a)) at a slightly lower energy ($E(0)=1.7809\,$eV) in
the absence of $B$ compared to the X$^{N=1}$ ($1.812\,$eV) in
Fig.1, which possibly gives rise to more extended localised
structure and a reduced threshold magnetic field. Spectral
splitting of the localised excitons in an anisotropic QR becomes
significant when $B\gtrsim4\,$T, where the Zeeman splitting
becomes comparable to the energy difference for perpendicular
linear polarisations ($\sim2\,$meV), similar to the case of
asymmetric quantum dots\cite{HtoonPRL}. The high ($E(\sigma^{-})$)
and low energy ($E(\sigma^{+})$) states of the exciton as a function of $B$ (Fig.4(b))
were obtained by Lorentzian fitting, whereby the excitonic $g$-factor
($g_{X}=[E(\sigma^{+})-E(\sigma^{-})]/\mu_{\rm{B}}B\sim-0.2$) and
diamagnetic coefficient
($\gamma=\frac{[E(\sigma^{+})+E(\sigma^{-})]/2-E(B=0)}{B^{2}}\sim1.3\,\mu$eV$T^{-2}$)
were also obtained.\\
\indent These small values can be attributed to the
small lateral area of the local structure\cite{Flatte}. After removing the Zeeman splitting and
diamagnetic shift (Fig.4(c)), an energy drift of $\sim 20\,\mu$eV is seen. However, it is difficult to define a significant
period. It is therefore possible that $14\,$T is an insufficient field to produce a phase
coherent delocalisation around the whole QR .
Nevertheless it would be interesting to try to induce delocalisation by either
applying a larger magnetic field or by suppressing the anisotropy in the QR.\\
\indent In conclusion, the presence of a strongly localised state
in an anisotropic QR was confirmed, where a crescent-like
asymmetric local structure gives rise to large polarisation
asymmetry for localised XX and X$^{N=2}$ and a small
diamagnetic coefficient for localised X$^{N=1}$. We
conclude a phase coherence around the whole rim is inhibited in the case of strong localisation, resulting in no significant excitonic AB oscillations up to $14\,$T.
\begin{acknowledgments}
This work was supported by the ``EC EuroMagNet II JRA8'' project, the European Commission from the 7th
framework programme ``Transnational Access", contract
N$^{\circ}$228043-EuroMagNet II-Integrated Activities, the NRF of
Korea Grant funded by MEST via NRF (2012R1A1A2006913 and
2010-0021173), BRL(2011-0001198), GRL program, and the KIST
institutional program.
\end{acknowledgments}


\begin{thebibliography}{}
\bibitem{Mano}%
T. Mano, T. Kuroda, K. Kuroda, and K. Sakoda, J. Nanophotonics
\textbf{3}, 031605 (2009).

\bibitem{Warburton}%
 R. J. Warburton, C. Schulhauser, D. Haft, C. Schaflein, K. Karrai, J. M. Garcia, W. Schoenfeld, and P. M. Petroff
 Phys. Rev. B \textbf{65}, 113303 (2002).

\bibitem{Spain}%
Benito Alen, Juan Martinez-Pastor, Daniel Granados, and Jorge M.
Garcia, Phys.Rev.B \textbf{72}, 155331 (2005).

\bibitem{Holland}%
N.A.J.M. Kleemans, I.M.A. Bominaar-Silkens, V. M. Fomin, V. N.
Gladilin, D. Gradados, A. G. Taboada, J. M. Carcia, P. Offermans,
U. Zeitler, P. C. M. Christianen, J. C. Maan, J. T. Devreese, and
P. M. Koenraad, Phys. Rev. Lett. \textbf{99}, 146808 (2007).

\bibitem{pcurrent}%
H. Bary-Soroker, O. Entin-Wohlman, and Y. Imry, Phys. Rev. Lett.
\textbf{101}, 057001 (2008).

\bibitem{ABnature}%
E. Bucks, R. Schuser, M. Heiblum, D. Mahalu, and V. Umansky,
Nature \textbf{391}, 871 (1998).

\bibitem{Teodoro}%
M.D. Teodoro, V.I. Campo Jr., V. Lopez-Richard, E. Marega Jr.,
G.E. Marques, Y. Galvao Gobato, F. Iikawa, M. J.S.P. Brasil, Z.Y.
AbuWaar, V.G.Dorogan, Y.I. Mazur, M. Benamara, and G.J. Salamo,
Phys. Rev. Lett. \textbf{90}, 186801 (2003).

\bibitem{Govorov1}%
A. O. Govorov, S. E. Ulloa, K. Karrai, R. J. Warburto,
 Phys. Rev. B \textbf{66}, 081309(R) (2002).

\bibitem{Govorov2}%
E. Ribeiro, A. O. Govorov, W. Carvalho, Jr., and G.
Medeiros-Ribeiro, Phys. Rev. Lett. \textbf{92}, 126402 (2004).

\bibitem{Ding}%
F. Ding, N. Akopian, B. Li, U. Perinetti, A. Govorov, F. M.
Peeters, C. C. Bofon, C. Deneke, Y. H. Chen, A. Rastelli, O. G.
Schmidt, and V. Zwiller, Phys. Rev. B \textbf{82}, 075309 (2010).

\bibitem{ZnTetypeII}%
R. Sellers, V. R. Whiteside, I. L. Kuskovsky, A. O. Govorov, and
B. D. McCombe, Phys. Rev. Lett. \textbf{100}, 136405 (2008).

\bibitem{Ulloa}%
Luis G. G. V. Dias de Silva, Sergio E. Ulloa, and Tigran V.
Shahbazyan, Phys. Rev. B \textbf{72}, 125327 (2005).

\bibitem{Taiwan}%
 T-C Lin, C-H Lin, H-S Ling, Y-J Fu, W-J Chang, S-D Lin, and C-P
 Lee, Phys. Rev. B \textbf{80}, 081304(R) (2009).

\bibitem{Kim}%
 H. D. Kim, K. Kyhm, R. A. Taylor, G. Nogues, K. C. Je, E. H. Lee and J. D. Song,
 Appl. Phys. Lett. \textbf{102}, 033112  (2013).

\bibitem{noAB}%
D. Haft, C. Schulhauser, A. O. Govorov, R. J. Warburton, K.
Karrai, J. M. Garcia, W. Schoenfeld, and P. M. Petroff, Physica
E \textbf{13}, 165 (2002).

\bibitem{HtoonPRL}%
 H. Htoon, S. A. Crooker, M. Furis, S. Jeong, Al. L. Efros, and V. I. Klimov,
 Phys. Rev. Lett. \textbf{102}, 017402 (2009).

\bibitem{Flatte}%
 J. van Bree, A. Yu. Silov, P. M. Koenraad, M. E. Flatte, and C. E. Pryor,
 Phys. Rev. B \textbf{85}, 165323 (2012).


\end{thebibliography}
\end{document}